# Using Artificial Intelligence to Improve Classroom Learning Experience

Shadeeb Hossain
Engineering Technology and
Information Sciences
DeVry University
New York , USA
[ORCID ID: 0000-0002-5224-7684]

*Abstract*—This paper explores advancements in Artificial Intelligence (AI) technologies to enhance classroom learning, highlighting contributions from companies like IBM, Microsoft, Google, and ChatGPT, as well as the potential of brain signal analysis. The focus is on improving students' learning experiences by using Machine Learning (ML) algorithms to (i) identify a student's preferred learning style (visual or auditory) and (ii) predict academic dropout risk. A Logistic Regression algorithm is applied for binary classification using six predictor variables, such as assessment scores, lesson duration, and preferred learning style, to accurately identify learning preferences. A case study, with 76,519 candidates and 35 predictor variables, assesses academic dropout risk using Logistic Regression, achieving a test accuracy of 87.39%. In comparison, the Stochastic Gradient Descent (SGD) classifier achieved an accuracy of 83.1% on the same dataset

## I. INTRODUCTION

Individual feedback to students and customized learning materials has a significant impact on their learning ability and have been areas of active research focus [1]. However, in the United States, due to the vast diversity in classroom populations, it becomes inherently difficult for educators to customize lessons and address individual students' problems [2]. Various factors contribute to the effectiveness of individual learning processes [3,4]. Questionnaires have often been used as a tool to predict an individual's learning style [5-8]. Learning analytics, which involves the collection, analysis, and use of data, has been suggested to improve students' learning experiences [9]. In most cases, these assessments have been used to generalize the overall learning patterns of a classroom rather than addressing the needs of individual students.

The concept of a SMART classroom incorporates both hardware and software components to adapt to dynamic learning patterns in a classroom, and it has been an area of ongoing research [10,11]. Aguilar et al. proposed that utilizing an institution's learning analytics, generated from a live SMART classroom environment, can produce more efficient results due to the instant real-time feedback provided [12]. The application of wireless sensors (WS) and the Internet of Things (IoT) for analyzing social and behavioral patterns in SMART classroom environments has been addressed across several testing platforms [13-21].

With evolving technology, it is now easier to predict an individual's learning profile and customize lessons accordingly. This personalization should eventually enhance a student's academic performance and motivation to learn.

Artificial Intelligence (AI) has gained popularity in the education sector due to its ability to (i) assess a student's learning needs, (ii) provide engaging assignments, (iii) automate grading, and (iv) assist students with homework assignments. A related and inseparable field in AI is the artificial neural network (ANN), which has been strongly inspired by biological neural networks. ANN involves predicting outputs based on input patterns [22], with applications ranging from classifying specific features of a tooth in scanned data to solving issues related to concentrator photovoltaics (CPV) technology [23,24].

One significant advantage of using ANN in a SMART classroom setting is its cognitive potential. The AI model should be able to classify a student's learning style as either (i) auditory, (ii) visual, or (iii) kinesthetic. It should also be able to assess a student's knowledge of a topic, categorizing them as either (i) beginner, (ii) intermediate, or (iii) advanced. The algorithm used in developing the ANN can help achieve the goal of personalized learning paces for students with different learning styles and abilities.

Lo et al. utilized a multi-layer feedforward neural network (MLFF) to develop a web-based learning system focused on students' cognitive styles [25]. Similarly, Curilem et al. proposed a mathematical model for an Intelligent Tutoring System (ITS) based on student behavior [26].

Beyond the cognitive capabilities of an ANN, the system offers predictive options, which are crucial for forecasting. Time series prediction using ANN has been previously applied in the financial industry [27-29] and in medical decision-making processes [30]. Likewise, in the education sector, ANN has been used for predicting school dropouts [27] and avoiding e-learning class evasion [32].

The proposed prototype, to be discussed shortly, can be easily implemented as a hardware module or integrated into a software algorithm. The proof-of-concept can also be developed into a mobile application, allowing students, educators, and guardians to access the learning trajectory in a SMART classroom.

The primary objective of this paper is to identify the influential factors of a SMART classroom that enhance a student's learning experience. Machine Learning (ML) algorithms are effective at identifying the variables that significantly impact a student's learning. This paper discusses two such instances: (i) identifying a student's preferred learning style (visual or auditory), as learning styles play a critical role in stimulating a student's engagement with lessons, and (ii) predicting variables that can assess whether a student is at academic risk of dropping out.

The article is divided into the following sections: (i) *State-of-the-art applications of AI in classroom teaching* – This section discusses the current evolution of teaching methodologies using technology. Tech giants such as Google, IBM, and Microsoft are key players in these revolutionary technologies. (ii) *The role of ChatGPT in higher education and student learning experiences* – This section explores how ChatGPT influences learning experiences in higher education. (iii) *Brain signal analysis* – This section focuses on the use of EEG and machine learning algorithms to predict attention fluctuations in classroom environments. It discusses both the potential and challenges of this technology in educational settings. (iv) *Proposed architecture for analyzing learning styles* – This section provides a detailed discussion of (a) the basic multi-layer ANN architecture and (b) the steps (including input parameters, hyperparameters, and expected outputs) for identifying the learning styles of individual students. (v) *Development of the algorithm* – This section covers the logistic regression algorithm for binary classification. (vi) *Case study on the classification of academic risk* – This section uses a dataset of 76,519 candidates with 35 predictor variables to determine if the target dependent variable indicates the likelihood to "graduate," "drop out," or remain "enrolled."

## II. STATE OF THE ART APPLICATION OF AI IN CLASSROOM TEACHING

Over the years, there have been multiple instances where technology has been integrated to improve the classroom learning experience. In 2019, NAO, an autonomous programmable robot, was used as a substitute for traditional educators to teach basic Chinese language to students in a classroom. The robot used emotion recognition and other complementary sensors to enhance student engagement during the learning process. The qualitative study was successful in terms of innovation and classroom engagement and could be considered an alternative to addressing the existing educator shortage.

Google Classroom is also integrating AI into its Google Workspace, Education Plus, allowing: (i) lessons to be transformed into interactive assignments with auto-grading tools, and (ii) students to receive real-time feedback on their assignments. The project is planned to launch as a beta test in selected schools and could potentially save teachers significant time in (i) grading assignments and (ii) providing insights into designing effective lesson plans. Additional features will offer students customized feedback and help them reach their target potential.

Another tech company, IBM, with its Watson Tutoring system, has partnered with Pearson to provide personalized learning experiences for students. It uses natural language processing, cognitive reasoning, and information retrieval to design its AI system. The system employs conversational chatbots to facilitate the exchange of information between the system and the student. It will analyze relevant data, including: (i) the time taken to respond to a question and (ii) the number of times the learning material has been reviewed or other similar indicators, to assess a student's learning pace and confidence.

Microsoft's AI-powered classrooms also harness technology to assist teachers and students in improving their speaking and math skills. Speaker Coach, a feature introduced by Microsoft in 2011, serves as an AI tool to help summarize presentation materials and highlight areas for improvement. Microsoft's AI math tools, such as Math Coach and Math Progress, allow teachers to input data to create new lessons and provide educators with insights into student progress.

Individual academic research institutions, such as DeNu, have proposed an AI technique called the Collaborative Logical Framework (CLF) that creates scenarios to support learning through interaction, collaboration, and discussion. The primary objectives of the framework are to (i) encourage students to work in groups and (ii) assist students in reaching agreements during conflicts when collaborating with other members.

The efforts of these organizations to enhance student learning experiences provide opportunities for (i) better student engagement in the classroom, (ii) individualized feedback, and (iii) catering to students' learning styles to help them reach their full potential.

## III. THE ROLE OF ChatGPT IN HIGHER EDUCATION AND STUDENT LEARNING EXPERIENCE

However, the most frequently used technological tool by both educators and students in their teaching-learning experience is ChatGPT. ChatGPT (Generative Pre-trained Transformer) was created by OpenAI and released to the public in March 2023. It uses data available until September 2021 to answer queries and employs reinforcement learning from human feedback to improve its responses.

Firat (2023) explored ChatGPT and other large language models (LLMs), such as Bard AI, as potential tools to transform the autodidactic learning experience [33]. ChatGPT and Bard can foster individual learning by providing customized solutions to students' queries. A study by Ausat et al. (2023) analyzed whether ChatGPT could replace the role of teachers in the classroom [34]. The study concluded that ChatGPT should be used as a tool to enhance the learning experience, rather than as a replacement for teachers.

Another study by Rasul et al. (2023) emphasized the need for caution when integrating ChatGPT into academics due to ethical concerns [35]. The research recommended adopting new assessment strategies, being cautious about false information, and including AI literacy as part of the graduate curriculum.

## IV. BRAIN SIGNAL ANALYSIS USING ANN (ARTIFICIAL NEURAL NETWORK) TO MONITOR CONCENTRATION OR ATTENTION SPAN IN CLASSROOM STUDENTS

In addition to the AI applications discussed earlier, it is crucial to study the impact of a student's inherent psychological and neurological characteristics when designing an efficient SMART classroom experience. Research shows that an average student's attention span declines approximately 10-15 minutes into a lecture.

Therefore, monitoring fluctuations in attention during class can provide insights for creating more effective lessons.

Ko et al. (2017) investigated electroencephalography (EEG) spectral changes in a classroom setting to study sustained attention [36]. Students were exposed to various visual stimuli, and their response times were analyzed to predict visual alertness. The power-frequency analysis demonstrated a relationship between EEG spectral dynamics and impaired performance in sustained attention tasks.

Another study by Chen et al. (2015) focused on recognizing students' attention levels using EEG signals in an e-learning environment [37]. The study aimed to develop an Attention-Aware System (AAS) by analyzing EEG data obtained during a continuous performance test. Machine learning algorithms, such as Support Vector Machines (SVM), were used to train and test the model. The results showed that the proposed AAS could accurately recognize a student's attention state—either 'high' or 'low'—with 89.25% accuracy.

The correlation between EEG data and students' attention spans can be applied in real-time classroom settings to gauge students' engagement with a particular lesson. Volunteering students could monitor brain signals in real time, and a decline in the overall class's concentration could prompt (i) a change in classroom activities or (ii) adjustments to future lessons to make them more engaging.

A Brain-Machine Interface (BMI) records electrical signals from the brain using neural probes. It employs CMOS technology to convert the analog signals from neurons into digital equivalents, which are then processed using Application-Specific Integrated Circuits (ASICs). The frequency of the received signals correlates with the brain's state. Abang et al. (2016) discuss the frequency bands and their corresponding brain states [38]. Frequency above 12 Hz are associated with an active and concentrated state, while lower frequencies (0.5 Hz-12 Hz) indicate a more relaxed state. An algorithm can process brain signals from participants and identify deteriorating attention when the brain wave frequency falls below 12 Hz.

Mehreen et al. (2019) used a combination of electroencephalography (EEG) electrodes, an accelerometer, and a gyroscope to detect drowsiness levels in subjects wearing sensors [39]. The extracted data, complemented by head movement and blink analysis, achieved an accuracy of 92%. ACI et al. (2019) developed algorithms using the Support Vector Machine (SVM) method to monitor the attention states of volunteers, achieving an average accuracy of 91.73% [40]. Niu et al. (2019) proposed a 1D convolutional neural network to capture frequency band information, predicting attention states with an accuracy of 96.4% [41].

Analyzing EEG reports with machine learning offers a feasible approach to evaluating students' concentration levels. However, there are several challenges associated with this technology, including: (i) discomfort from wearing the equipment, (ii) high costs, (iii) the need for technical expertise during initial evaluation, and (iv) potential false reporting, as students may alter their behavior knowing they are being monitored.

The following section proposes a prototype architecture that integrates machine learning to identify a student's preferred learning style, whether (i) visual or (ii) auditory. The system could also incorporate additional features such as (i) tracking lesson progress and (ii) assessing the "level of prior knowledge" on a topic, whether beginner or advanced, for an enhanced learning experience.

## V. PROPOSED ARCHITECTURE FOR AI CLASSROOM TEACHING USING FEEDBACK ANN

*A. The Basic Multi-Layer ANN Architecture*

The architecture of an Artificial Neural Network (ANN) primarily consists of three layers: (1) Input Layer, (2) Hidden Layer, and (3) Output Layer. **Fig. 1** shows a simple architecture for a multi-layered ANN. The input layer receives the necessary data from the external environment. The parameters of interest include: (i) responses to assessment questions based on visual and auditory learning styles, (ii) time required to comprehend the learning material, (iii) the student's preferred learning style from past experiences, (iv) time of day, (v) instructor score (how students assess the teaching style of the educator), and (vi) duration of the lesson.

The hidden layer is responsible for processing the data before passing it to the next architecture level. This involves assigning weights to each input parameter. Weights can be assigned equally to all parameters, or certain parameters may be given higher weights than others (for example, responses to assessment questions might have a higher weight compared to the instructor score).

The output can be a binary classification, where auditory learners are classified as binary '0' and visual learners as binary '1'. A similar approach can be used to classify (i) advanced learners, (ii) intermediate learners, and (iii) beginners. This is an example of supervised learning, and logistic regression can be employed for binary classification. Based on the output generated by the algorithm, students can be provided with resource materials that correspond to their preferred learning style.

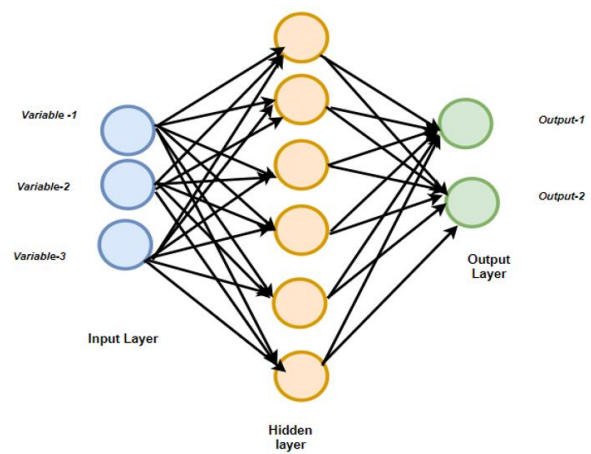

**Fig.1:** The schematic architecture of a Multi-layer ANN developed for " SMART Classroom Learning".

*B. Identifying Learning Style of Individual Student*

Most teaching materials are generic, yet classrooms are typically diverse, with students exhibiting different learning styles. This diversity presents a challenge for educators, as they strive to ensure full student engagement in the learning process. The use of Artificial Neural Networks (ANN) can effectively identify a struggling student's inherent learning style and provide personalized resources aligned with their preferences. Additionally, educators can receive real-time updates on declining concentration levels within their classrooms. Both students and educators can access learning analytics in SMART classrooms, improving the overall quality of the learning experience.

**Fig. 2(a)** presents the schematic of the proposed architecture for a SMART classroom environment with students of varying learning abilities. **Fig. 2(b)** illustrates how the system uses AI to identify each student's learning style and provide them with the most effective study resources.

The Felder-Silverman Learning Style Model (FSLSM) can identify an individual's preferred learning style. Visual learners favor graphs, charts, and diagrams as their learning resources, while auditory learners prefer activities centered on speaking and listening [42, 43].

for binary classification, with the output categorized as either '0' for an auditory learner or '1' for a visual learner. If the ratio of the **visual score** (generated from the tally of high scores when presented with visually stimulating study material) to the total score is above 65%, the student is considered a visual learner, and vice versa.

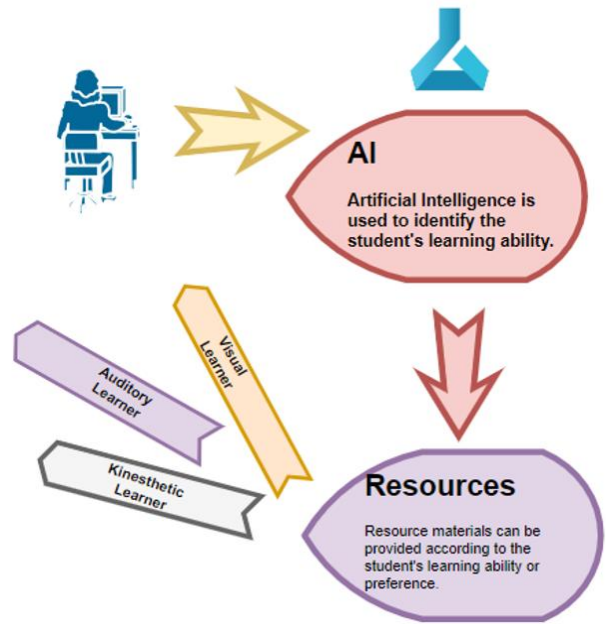

**Fig. 2(b):** Simplified schematic of using ANN to identify the student's initial learning style

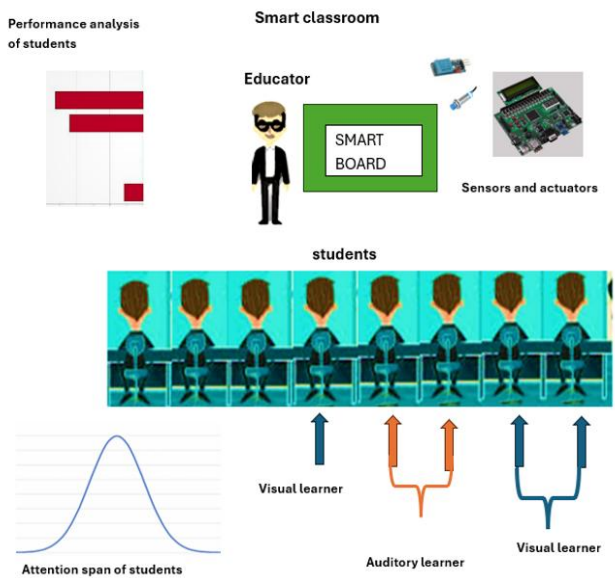

**Fig. 2(a):** Typical classroom environment with students with different learning abilities attending the same lecture.

**Fig. 3(a)** shows a flowchart of the steps that can be implemented to identify the learning style of individual students. In the initial steps, a topic of interest is chosen, and the student is presented with study materials on that topic in the form of charts, graphs, essays, or other visually stimulating resources. A quantitative assessment is then performed to evaluate the student's understanding of the topic. The process is repeated for the same student, using the same learning topic (but a different subsection to avoid overlap), with resources designed for auditory learners. A quantitative assessment is performed again to monitor the student's response to auditory stimuli.

In **Fig. 3(b)**, the tally for the scores generated from visual and auditory stimulating materials is shown. Logistic regression with a sigmoid activation function can be applied

The proposed model for analyzing a student's learning style, as shown in **Fig. 3**, can be easily implemented in a software module using the Python programming language. For a more advanced system, the Tobii eye-tracking system, as proposed by Mehigan et al., can be integrated with our simple model in **Fig. 3** to analyze gaze patterns, heat maps, and fixation counts, thereby identifying visual and auditory learners' concentration in real-time [44]. This biometric technology is ideal for both e-learning and high-tech SMART classroom environments. However, integrating biometrics into the system presents challenges, as it requires participant consent each time and may cause student discomfort due to the associated gadgets and constant monitoring.

According to Joko Stuarto, tutor commitment, lesson delivery, and lesson planning account for 24.84% and 38.69%, respectively, in the effectiveness of the learning process [3]. Therefore, it is important that other input parameters, as highlighted earlier in this section, are considered when developing an AI algorithm to identify the learning style of individual students. In addition to the factor of 'response to assessment questions,' other factors, such as 'instructor score,' can be critical in the decision-making process.

Figure 4 illustrates a digital logic equivalent of the hidden layer in the proposed neural network. In the model, a student labeled as an auditory learner (Student A) is evaluated across different instructors (Teacher 1, Teacher 2, and so on) to reconfirm their learning style. This ensures that other input variables, such as "instructor score," are incorporated into the logistic regression model. The model also includes input parameters like "time for delivery of instruction," labeled as "output style for Day 1, Teacher 1."

be a visual learner, the binary classification is '1,' and if not, it is '0.'

Rahman et al. highlighted that recalling or revising past learning on a topic can enhance the learning experience for new, relevant information [4]. Assessments can be conducted to gauge a student's prior knowledge of a topic before introducing new material to avoid overwhelming them.

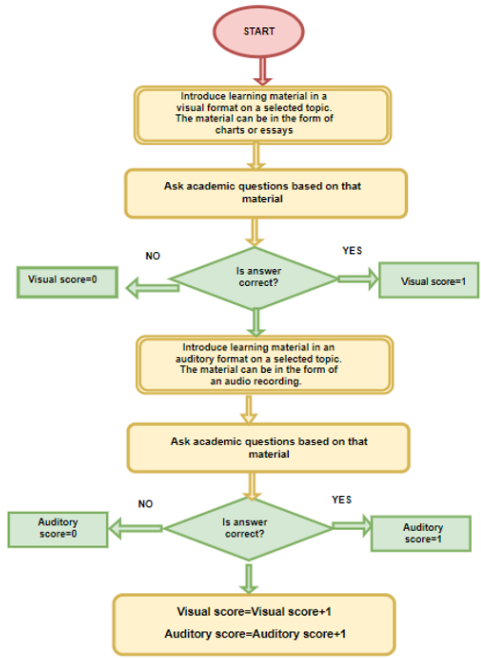

**Fig. 3(a):** Flowchart for determining learning style by presenting visual and auditory based materials.

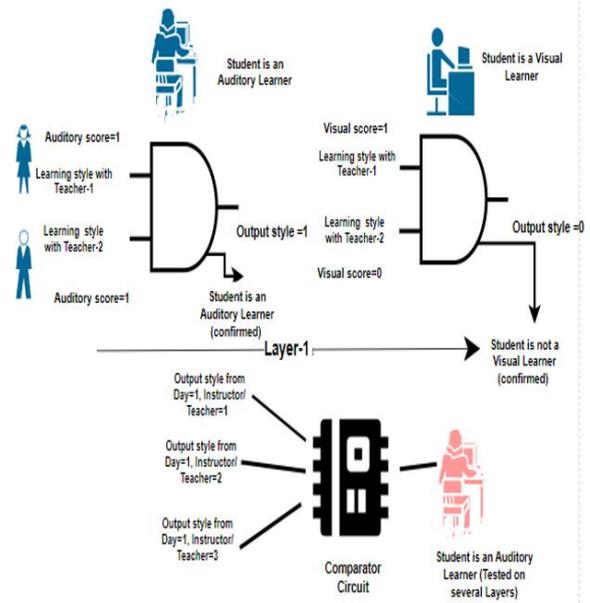

**Fig. 4:** Digital Logic Equivalent of Hidden Layers in the ANN.

Figure 5 illustrates the steps to evaluate a student's understanding of a lesson. The process begins with an initial assessment that includes questions to evaluate the student's knowledge of the topic. If the student achieves a high score on the quantitative assessment, they can then be tested with more advanced questions that require deeper knowledge and critical thinking. The results are used to determine whether the student has a "beginner" or "advanced" understanding of the subject.

Figure 5(b) shows how the process can be further enhanced by including the combined scores of the entire class to compare their overall "beginner" or "advanced" understanding of the topic. This information can help the instructor decide how to proceed with the lesson. The advantage of using a smart classroom and machine learning is that instruction becomes dynamic, as assessments are performed in real time, allowing the necessary adjustments to accommodate students' learning preferences.

An existing database of teaching resources tailored for both "beginner" and "advanced" students can also be utilized. This helps address challenges such as (i) large classroom sizes, (ii) classroom management, and (iii) student-teacher ratios.

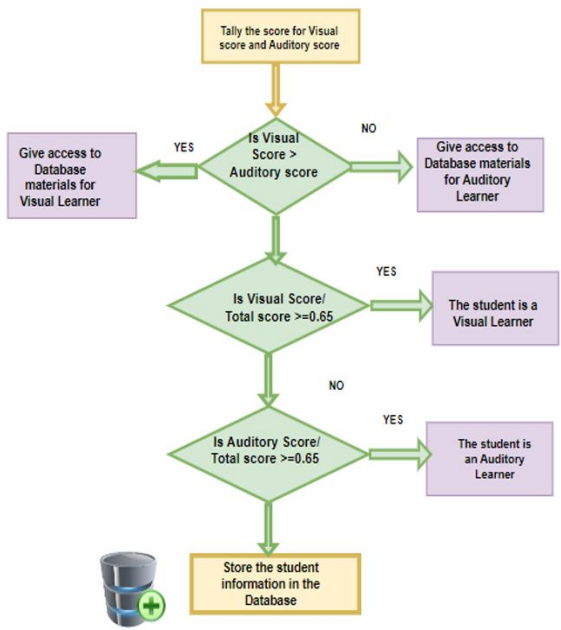

**Fig. 3(b):** Flowchart for determining learning style and storing information in the database for future reference.

A final decision on whether the student is a visual or auditory learner is made by considering all input parameters and the correlation coefficient. If the student is determined to

## VI. DEVELOPMENT OF THE ALGORITHM

The time-dependent mathematical expression for the input and output layers includes the following parameters, which are influenced by the Zel-Andreas mathematical model

[45]: activation, threshold value, activation function, and output function.

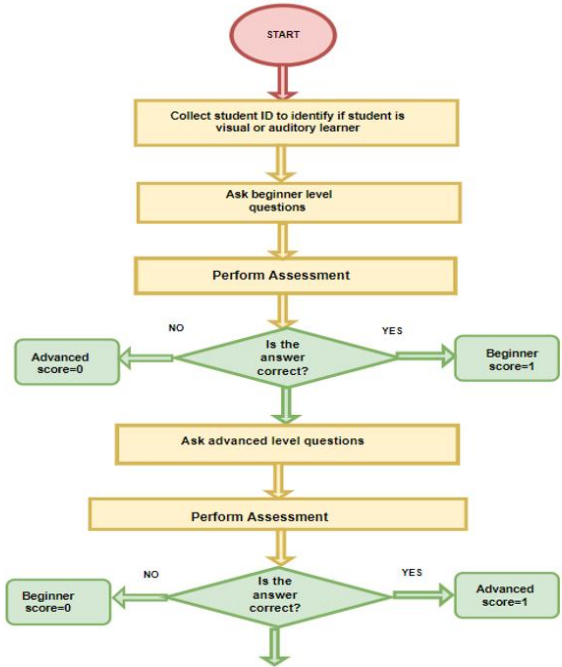

**Fig.5 (a):** Flowchart for predicting Learning Stage : Beginner or Advanced.

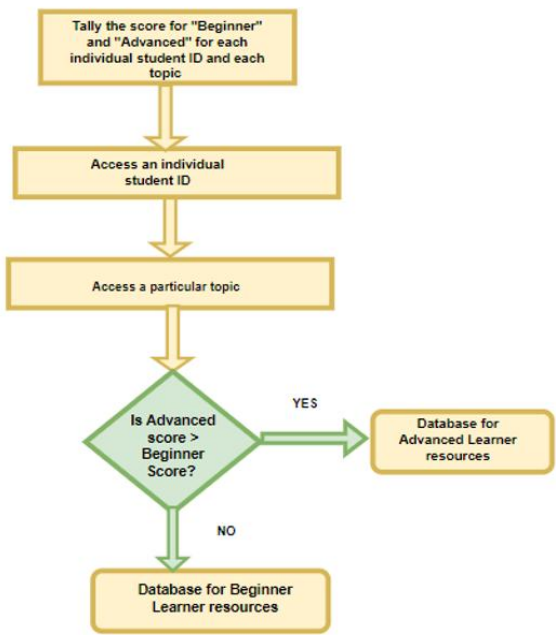

**Fig.5 (b):** Flowchart for directing to "Beginner" or "Advanced" Learner resources.

There are also various input variables used in this multiple or polynomial regression model. These variables include: (i) assessment question scores, (ii) time taken to comprehend a lesson, (iii) preferred learning style (which may differ from the one determined by the model), (iv) time of lesson delivery, and (v) instructor score.

The regression model can be expressed as dependent variable 'y' ( which is either visual or auditory) and independent input variables, $x_{i-N}$, where N is the number of input parameters = 6.
Mathematically the multiple and polynomial regression models can be expressed as equations (1) and (2) respectively. [46-50]:

$$y = \beta_0 + \beta_1 x_1 + \beta_2 x_2 ---- \beta_N x_N \quad (1)$$

$$y = \beta_0 + \beta_1 x_1 + \beta_2 x_2^2 ------ \beta_N x_N^N \quad (2)$$

where, $\beta$ represents the correlation coefficients for each of the input parameters.

The least square regression (LSR) methos can be used to find the best fit line or curve for the parameters which minimizes the error [51,52]. Mathematically, it can be represented as equation (3).

$$LSR = min \sum_{i=1}^{N}(y_{actual} - (\widehat{y_{predicted}}))^2 \quad (3)$$

where, LSR is the least square regression output and $y_{predicted}$ is the output calculated from equation (1) or (2) respectively.
To compute the predicted learning style, $\widehat{y_{predicted}}$, each attribute can be accounted at a time and can be represented by equation (4).

$$\widehat{y_{predicted}} = \beta_{0,n} + \beta_{1,n} x_n, \text{ where } n|0 \le n < N \quad (4)$$

In LSR, the objective is to minimize the error between the actual and predicted value. Mathematically it can be represented with the concept applied in [50].

$$\frac{\partial}{\partial \beta_{0,n}} \sum_{i=1}^{N}((y - (\beta_{0,n} + \beta_{1,n} x_n))^2 = 0 \quad (5)$$

$$\frac{\partial}{\partial \beta_{1,n}} \sum_{i=1}^{N}((y - (\beta_{0,n} + \beta_{1,n} x_n))^2 = 0 \quad (6)$$

$$LSR_N^2 = \frac{|S_{xn,y}|^2}{|S_{xn}^2| S_y^2} \quad (7)$$

where, $S_{xn,y}$ is the cross co-variance and $S_{xn}$ and $S_y$ are the auto-covariance respectively.

The value of the coefficients for each input variable can be calculated using equation (8) and (9).

$$\beta_{1,N} = \frac{S_{yx}}{S_{xx}} = \frac{\sum_{i=1}^{N}(y_i - \bar{y})(x_i - \bar{x})}{\sum_{i=1}^{N}(x_i - \bar{x})^2} \quad (8)$$

$$\beta_{0,N} = \bar{y} - \beta_{1,N} \overline{x_n} \quad (9)$$

*A. Using Logistic Regression for Binary Classification*

The logistic function uses an S-shaped curve that takes a real value number and converts to a value between '0' and '1'. Mathematically the logistic regression equation can be expressed as equation (10).

$$y = \frac{e^{(\beta_0+\beta_1 x_1+\beta_2 x_2+\beta_3 x_3+\beta_4 x_4+\beta_5 x_5+\beta_6 x_6)}}{1+e^{(\beta_0+\beta_1 x_1+\beta_2 x_2+\beta_3 x_3+\beta_4 x_4+\beta_5 x_5+\beta_6 x_6)}} \quad (10)$$

To perform logistic regression, the following steps are implemented: (i) data preparation, (ii) identifying features and target variables, (iii) splitting the data into training and testing sets, (iv) standardization, (v) creating a logistic regression model, (vi) model evaluation, and (vii) prediction. For data preparation, students can be provided with differentiated learning resources, and their assessment scores in each scenario can be recorded. Simultaneously, other independent variables, such as the time taken to comprehend the learning resources, time of instruction delivery, and instruction style, may also need to be recorded. The six independent variables listed in the section "Development of the Algorithm" are the features, while the target variable is the 'learning style.'

The dataset can then be split into training and testing sets, and the feature variables are standardized to have a mean of 0 and a standard deviation of 1 using the '*StandardScaler*' algorithm.

The logistic regression model can be used to allow the model to learn the relationship between input features and output labels. The training process involves optimization algorithms, such as gradient descent, which adjust the model's weights to best fit the training data and reduce the loss. Once trained, the model can be evaluated for accuracy and used to make predictions about unknown students based on their input performance.

Multiple linear regression often faces the problem of multicollinearity between predictor variables (e.g., collinearity between time of instruction and instructor score), which can be minimized using Lasso and Ridge regularization methods. Other hyperparameters can also be adjusted simultaneously to reduce the mean squared error (MSE) and improve prediction accuracy.

VII. CASE STUDY ON CLASSIFICATION OF ACADEMIC RISK

The dataset from Reade et al. (2024) is used as a case study to analyze how machine learning algorithms can aid in effective classification, where predictor variables can determine if a candidate is likely to: (i) "graduate," (ii) "drop out," or (iii) remain "enrolled" [51]. Approximately 35 predictor variables were used in the dataset, and the number of candidates included in the study was 76,519.

Factors such as whether evening or daytime classes were attended, parents' academic qualifications, student demographics (such as age and gender), scholarship status, employment status while enrolled, and the degree program enrolled are all considered in training the model.

The training and test datasets were split into a 70:30 ratio, and a logistic regression model was used for training. The maximum number of iterations was limited to 1000, and the Limited-memory Broyden-Fletcher-Goldfarb-Shanno (L-BFGS) algorithm was used as the solver. Fig. 6 shows the statistics of the accuracy, precision and F1 score obtained from logistic regression model. The model obtained a training accuracy of 87.85% and test accuracy of 87.39%.

```
Class distribution in the training data:
Target
Graduate    36282
Dropout     25296
Enrolled    14940
Target          1
Name: count, dtype: int64
Training Accuracy: 87.85%
Training Precision: 85.92%
Training Recall: 90.50%
Training F1 Score: 88.15%
Test Accuracy: 87.39%
Test Precision: 85.75%
Test Recall: 89.76%
Test F1 Score: 87.71%
```

**Fig. 6 :** shows the statistics of the accuracy, precision and F1 score obtained from logistic regression model.

Stochastic Gradient Descent (SGD) classifier can also be used with this dataset as they can handle larger datasets with complex models. SGD classifier is a linear classifier that uses stochastic gradient descent for optimization. 'Log' loss corresponding to logistic regression loss function and a constant learning rate of 0.01 was used. The training and test dataset was split into 70:30 ratio and the number of epochs was limited to 100. As shown in Fig.7 the training accuracy was 84.3% and the test accuracy was 83.1%.

The accuracy of this model can be further improved by using (i) LASSO regression to identify multicollinearity or (ii) using L2 regularization to reduce overfitting (iii) tuning hyperparameters such as 'max iteration', tolerance (tol'), alpha regularization.

VIII. CONCLUSION

This paper provides a comprehensive review of advancements in AI within the education industry. It discusses state-of-the-art AI teaching applications and the contributions of leading tech giants such as IBM, Google, Microsoft, and others in enhancing the learning experience. Additionally, the role of ChatGPT is examined, along with a detailed description of proposed architecture designed to study and determine the learning abilities of diverse student classrooms.

Six predictor variables(i) assessment question scores, (ii) time taken to comprehend a lesson, (iii) preferred learning style (which may differ from the model's determination), (iv) time of lesson delivery, and (v) instructor score—are used to determine the target dependent variable, whether a student is a "Visual" or "Auditory" learner. Linear logistic regression is proposed to train the model for binary classification, with the model's accuracy tunable through hyperparameter adjustments.

Another case study on academic risk classification is also reviewed. The dataset, sourced from Kaggle, contains 76,519 candidates and includes 35 predictor variables to determine the target outcome: whether a candidate is likely to "graduate," "drop out," or remain "enrolled." When the Stochastic Gradient Descent (SGD) classifier was applied to this dataset, the test accuracy was only 83.1%. However, using logistic regression with the L-BFGS algorithm as the solver, a test accuracy of 87.39% was achieved.

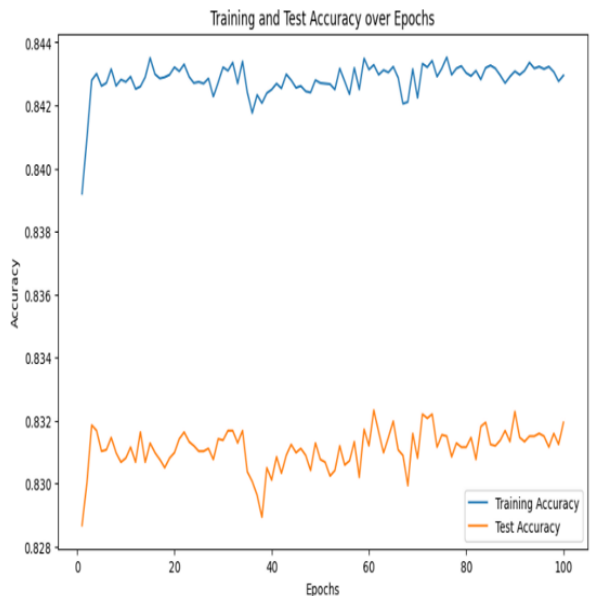

**Fig.7** : Training and Test accuracy obtained from Stochastic Gradient Descent Classifier.